\input ./style/arxiv-general.cfg
\documentclass[aoas,MSNbibl,nameyear,seceqn,dvips]{arximspdf}
\makeatletter
   \@ifpackageloaded{graphicx}{}{\usepackage{graphicx}}
\makeatother
\usepackage{dcolumn}

\doi{10.1214/15-AOAS856}% Updated by VTEXPTS2LaTeX.exe, 21.08.2015 08:32
\volume{9}
\issue{3}
\pubyear{2015}
\firstpage{1433}
\lastpage{1458}
\docsubty{FLA}

\makeatletter
\newcolumntype{d}[1]{D{.}{.}{#1}}

\newtheorem{prop}{Proposition}[section]
\makeatother

\begin{document}
\begin{frontmatter}

%\dochead{}
\title{A new set of asymmetric filters for tracking the short-term trend in real-time}
\runtitle{Tracking the short-term trend in real time}

\begin{aug}
% Corresponding author: Estela Dagum - estela.beedagum@unibo.it% Updated by VTEXPTS2LaTeX.exe, 21.08.2015 12:45
%Updated by VTEXPTS2LaTeX.exe, 21.08.2015 08:32
\author[A]{\fnms{Estela} \snm{Bee Dagum}\corref{}\ead[label=e1,mark]{estela.beedagum@unibo.it}}
\and
\author[A]{\fnms{Silvia} \snm{Bianconcini}\ead[label=e2]{silvia.bianconcini@unibo.it}}
\runauthor{E.~B. Dagum and S. Bianconcini}
\address[A]{Department of Statistical Sciences\\
University of Bologna\\
Via Belle Arti, 41 - 40126 Bologna\\
Italy\\
\printead{e1}}
\affiliation{University of Bologna}
%\runauthor{}
%\dedicated{}
\end{aug}

% HISTORY:
%
\received{\smonth{9} \syear{2014}}% Updated by VTEXPTS2LaTeX.exe,
%21.08.2015 08:32
%
\revised{\smonth{6} \syear{2015}}% Updated by VTEXPTS2LaTeX.exe,
%21.08.2015 08:32

% ABSTRACT
%
\begin{abstract}
For assessing in real time the short-term trend of major economic
indicators, official statistical agencies generally rely on asymmetric
filters that were developed by Musgrave in 1964. However, the use of
the latter introduces revisions as new observations are added to the
series and, from a policy-making viewpoint, they are too slow in
detecting true turning points. In this paper, we use a reproducing
kernel methodology to derive asymmetric filters that converge quickly
and monotonically to the corresponding symmetric one.
We show theoretically that proposed criteria for time-varying bandwidth
selection produce real-time trend-cycle filters to be preferred to the
Musgrave filters from the viewpoint of revisions and time delay to
detect true turning points.
We use a set of leading, coincident and lagging indicators of the US
economy to illustrate the potential gains statistical agencies could
have by also using our methods in their practice.
\end{abstract}

% KEYWORDS
% Pirmas kwd is didziosios raides
%
\end{frontmatter}

%s1 #&#
\section{Introduction}\label{sec1}

In recent years, statistical agencies have shown an interest in
providing trend-cycle or smoothed seasonally adjusted graphs to
evaluate the stage of the cycle at which the economy stands. This is
known as recession and recovery analysis, and differs from business
cycle studies where cyclical fluctuations are measured around a
long-term trend to estimate complete business cycles [see, e.g.,
\citet{Hod97,Chr03,Aze06,Aze11,deC12,deC14}]. Among other
reasons, this interest originated from the recent crisis and major
economic and financial changes of global nature which have introduced
more variability in the data. The US entered in recession in December
2007 till June 2009, and this has produced a chain reaction all over
the world. There is no evidence of a fast recovery as in previous
recessions. The economic growth is sluggish and high levels of
unemployment have been observed. It has become difficult to determine
the direction of the short-term trend by simply looking at month to
month (quarter to quarter) changes of seasonally adjusted values,
particularly to assess the upcoming of a true turning point. Failure in
providing reliable trend-cycle estimates in real time could lead to the
adoption of counteract policies that will affect the whole economy in a
negative way.

The linear filter developed by \citet{Hen16} is one of the most
frequently applied to estimate the trend-cycle component of seasonally
adjusted economic indicators. It is available in nonparametric seasonal
adjustment software, such as the US Bureau of the Census X11 method
[\citet{Shi67}] and its variants, X11/X12ARIMA and X13. The Henderson
smoother has the property that fitted to exact cubic functions will
reproduce their values, and fitted to stochastic cubic polynomials it
will give smoother results than those estimated by ordinary least
squares. The properties and limitations of the Henderson filters have
been extensively discussed by many authors, among them, \citet{Cho81},
\citet{KenDur82}, \citet{DagLan87}, \citet{Dag96}, \citet{GraTho96}, \citet
{Loa99}, \citet{LadQue01}, \citet{FinMar06}, \citeauthor{DagLua09a} (\citeyear{DagLua09a,DagLua12}).
\citet{DagBia08} represented the Henderson filter using
Reproducing Kernel Hilbert Space (RKHS) methodology [we refer the
reader to \citet{Ber03} for a detail description of RKHS]. Their
approach is based on a theoretical result due to \citet{Ber93},
according to which a kernel estimator of order $p$ can always be
decomposed into the product of a reproducing kernel $R_{p-1}$,
belonging to the space of polynomials of degree at most $p-1$, and a
probability density function $f_{0}$ with finite moments up to order
$2p$. The authors found that a kernel function obtained as the product
of the biweight density function and the sum of its orthonormal
polynomials is particularly suitable when the length of the filter is
rather short, say, between 5 to 23 terms, which are those often applied
by statistical agencies.

At the beginning and end of the sample period, the Henderson filter of
length, say, $2m+1$, cannot be applied to the $m$ data points, hence,
only asymmetric filters can be used. The estimates of the real time
trend are then subject to revisions due to the innovations brought by
the new data entering in the estimation and to the fact that the
asymmetric filters are time varying in the sense of being different for
each of the $m$ data points.

In this paper, we propose a new set of asymmetric weights to replace
the Musgrave ones officially adopted by statistical agencies to detect
the direction of the short-term trend in real time. From an applied
viewpoint, we are motivated by the need of obtaining reliable
short-term estimates in real time, which can be more useful from a
policy-making viewpoint. We apply the new filters to leading,
coincident and lagging indicators of the US economy, which is known to
be a key player from an international macroeconomic perspective. We
will concentrate on the reduction of revisions only due to filter
changes, and ignore those introduced by new innovations entered with
new data. In other words, the filter revisions depend on how close the
asymmetric filters are with respect to the symmetric one [\citet
{DagLan87,Dag96}]. Besides the filter revisions, we shall deal with the
time delay to identify the upcoming of a true turning point. Another
important property analyzed for the new set of asymmetric filters is
the time path followed by the last trend-cycle point as new
observations are added to the series. This is obtained by calculating
the number of months (quarters) it takes for the last trend-cycle
estimate to identify a true turning point in the same position of the
final trend-cycle data. An optimal asymmetric filter should have a time
path that converges fast and monotonically to the final estimate as new
observations are added to the series.

Several authors have studied the properties and limitations of the
Musgrave filters [\citet
{Lan85,Doh01,GraTho02,QueLad03},
\citeauthor{DagLua09b} (\citeyear{DagLua09b,DagLua12}),
\citet{BiaQue10}].
\citeauthor{DagBia08} (\citeyear{DagBia08,DagBia13}) introduced
a RKHS representation of the asymmetric
filters of \citet{Mus64}. In the RKHS framework, given the density
function (in our case the biweight), once the length of the symmetric
filter is chosen, say $2m+1$, the statistical properties of the
asymmetric filters are strongly affected by the bandwidth parameter of
the kernel function from which the weights are derived. In previous
works, \citeauthor{DagBia08} (\citeyear{DagBia08,DagBia13}) made the bandwidth parameters equal for
all the asymmetric filters (global time-invariant bandwidth) to closely
approximate the Musgrave filters.

Additionally, we propose here time varying bandwidth parameters since
the asymmetric filters are time varying. We consider three specific
criteria of bandwidth selection based on the minimization of the following:
\begin{longlist}[1.]
\item[1.] the distance between the transfer functions of asymmetric and
symmetric filters,
\item[2.] the distance between the gain functions of asymmetric and
symmetric filters, and
\item[3.] the phase shift function over the domain of the signal.
\end{longlist}

Section~\ref{sec2} presents a motivating example using the US New Orders for
Durable Goods (NODG) series. Section~\ref{sec3} gives the RKHS representations
of the Henderson and Musgrave linear filters, and discusses the
discretization of the continuous kernel functions when applied to data.
Section~\ref{sec4} deals with the time-varying optimal bandwidth selection where
a filter is defined as optimal if: (1) it minimizes the revisions
between last point and final trend-cycle values as new observations are
added, and (2) reduces the time delay to signal the upcoming of a true
turning point. Section~\ref{sec5} provides an empirical application to leading,
coincident and lagging indicators of the US economy. Finally, Section~\ref{sec6}
gives the conclusions.

%s2 #&#
\section{Motivating example: US new orders for durable goods}\label{sec2}
\begin{figure}[b]

\includegraphics{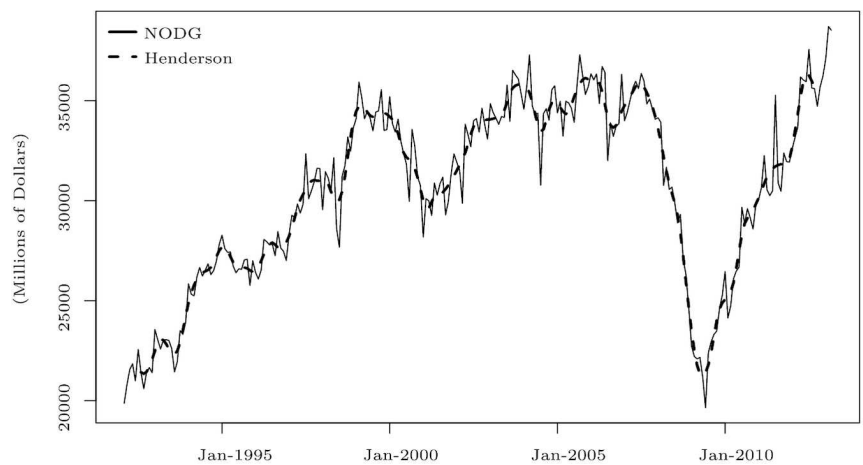}

\caption{New Orders for Durable Goods, US: seasonally adjusted series
and trend-cycle estimates obtained with the 13-term Henderson filter.
Source: US Census Bureau.}\label{NODG}
\end{figure}

The monthly series of US New Orders for Durable Goods (NODG), published
by the US Census Bureau, measures the volume of orders of goods whose
intended lifespan is three years or more. Approximately 60 percent of
the orders are for cars and trucks, with building materials, furniture
and household items accounting for most of the remaining part. The NODG
series is a leading indicator of US manufacturing activity, and an
increase in orders is considered as more future business for
manufacturers. The market often moves on accordingly in spite of its
high volatility, hence, it represents an important indicator of the
state of the economy, allowing to detect shifts in the US economy up to
six months in advance. Figure~\ref{NODG} illustrates the final vintage
data of the monthly NODG series for the period February 1992--March
2013. It is evident that the NODG peaked in the middle of 2007, and
underwent thenceforth a very steep decline up to June 2009, that has
been identified by the Business Cycle Dating Committee of the National
Bureau of Economic Research (NBER) to be the last trough in the US economy.
The dashed line overlaid to the seasonally adjusted NODG series in
Figure~\ref{NODG} is the nonparametric estimate of the corresponding
trend-cycle component produced by the application of the 13-term
symmetric filter due to \citet{Hen16}.

%f1 #&#

\begin{figure}

\includegraphics{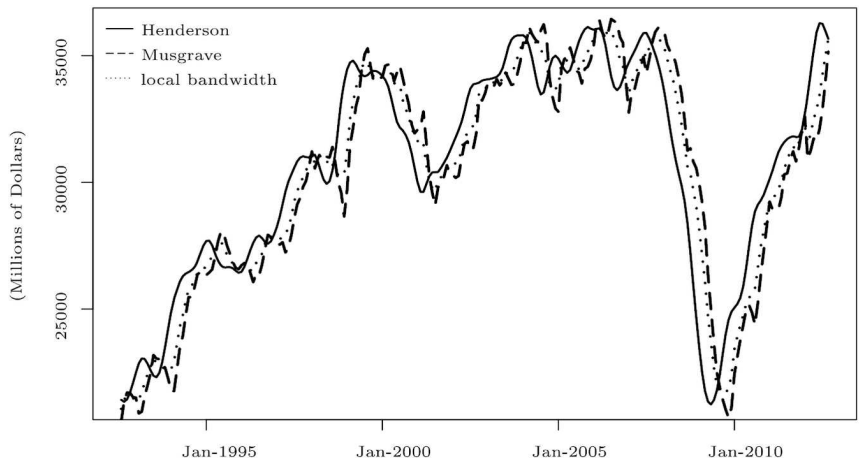}

\caption{New Orders for Durable Goods, US: trend-cycle estimates based
on Henderson filter, last point Musgrave and RKHS asymmetric filters,
respectively.}\label{NODGrev}
\end{figure}

It is evident from Figure~\ref{NODG} that the two-sided estimates of
the signal are not available for the first and last six months, the
latter being the most important for short-term trend prediction. The
corresponding estimates are derived using asymmetric filters due to
\citet{Mus64}. They are known to possess the good property of fast
detection of turning points, but they tend to introduce large revisions
when new observations are added to the series. This is illustrated in
Figure~\ref{NODGrev} for the last point Musgrave filter that is the
most important since it provides the real time trend-cycle estimate
corresponding to the current observation.
Besides the phase shift effect typical of asymmetric filters, that
produces a temporal displacement of the point of maxima and minima of
the input series, a crude measure of the size of the total revision of
the asymmetric filter is given by the distance, for each point in time,
between the estimate obtained by its application (long dash line) and
the final estimate derived by using the symmetric filter (solid line).

%f2 #&#

To overcome the main limitations of the Musgrave filters, \citet
{DagBia08} have provided an equivalent kernel representation of the
symmetric Henderson filter and derived the corresponding asymmetric
filters using the Reproducing Kernel Hilbert Space (RKHS) methodology.
The main advantage of the asymmetric kernel filters with respect to the
Musgrave ones is that the former are derived following the same
criteria as the symmetric filter, whereas the latter are determined
based on different optimization criteria. Having chosen the length of
the filter, the properties of these asymmetric kernels are strongly
dependent on bandwidth parameters. The current authors originally made
the bandwidth parameter equal for all the asymmetric filters (global
time-invariant bandwidth) to closely approximate the Musgrave filters
and ensure a fast convergence to the corresponding symmetric one.

\begin{figure}[b]

\includegraphics{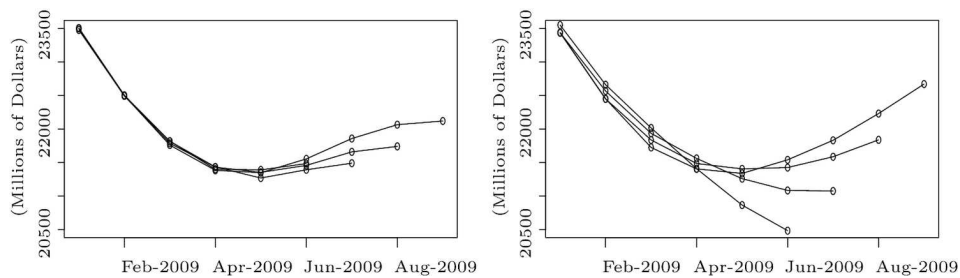}

\caption{US NODG series. Revision path of the June 2009 (turning point)
estimate as one observation is added at a time up to December 2009
(final estimate) using the optimal last point asymmetric kernel (left)
and Musgrave (right) filters, respectively.}\label{porc}
\end{figure}

In this paper, several criteria for bandwidth selection are proposed
based on specific properties that the corresponding asymmetric filters
should satisfy. As a specific case, Figure~\ref{NODGrev} shows the
behavior of the last point kernel filter whose bandwidth parameter has
been selected in order to ensure more accurate predictions. In
particular, the latter has been chosen to minimize the distance between
the gain functions of the last point and symmetric kernel filters. It
can be noticed that over the whole sample span, the kernel filter
(dotted line) is the closest to the final estimates (solid line). As
discussed in the subsequent sections, the revisions are almost 50
percent smaller than those introduced by the Musgrave filter.
However, it should be noticed that a reduction in the revisions does
not necessarily imply a reduction in the time lag to signal the
upcoming of a true turning point. This is obtained by calculating the
number of months it takes for the revised real time trend-cycle to
signal a turning point in the same position as in the final trend-cycle
series. For the June 2009 turning point observed for the NODG series,
this is illustrated in Figure~\ref{porc} for the last point Musgrave
(right) and optimal kernel (left) filters. This figure gives the
revision path of the last available point (June 2009) as we keep adding
one observation at a time up to December 2009, when the final estimate
is achieved. It can be noticed that after adding only one month to the
series ending in June, the turning point is clearly detected by the
RKHS filter, whereas two months are required by the Musgrave filter.
%
%f3 #&#

Full details will be given in the sequel; it suffices to say at this
point that the set of asymmetric filters for detecting the short-term
trend in real time introduced in this study provides better estimates
than the classically applied Musgrave filters. The improvements are
reflected in the size of total revisions and time delay to identify the
upcoming of a true turning point. This is illustrated more extensively
in Section~\ref{sec5}.

%s3 #&#
\section{Linear filters in RKHS}\label{sec3}

Let $\{y_{t}, t=1,\ldots,N\}$ denote the input series, supposed to be
seasonally adjusted where trading day variations and extreme values, if
present, have been also removed. We assume that it can be decomposed
into the sum of a systematic component (signal) $g_{t}$, that
represents the trend and cycle usually estimated jointly, plus an
erratic component $u_{t}$, called the noise, such that
%
%e3.1 #&#
\begin{equation}
y_{t}=g_{t}+u_{t}. \label{1}
\end{equation}
The noise $u_{t}$ can be either a white noise, $\operatorname{WN}(0,\sigma_{u}^{2})$,
or, more generally, a stationary and invertible
AutoRegressive Moving Average (ARMA) process.
On the other hand, the signal $g_{t}, t=1, \ldots, T$, is assumed to
be a smooth function of time, such that it can be represented \emph
{locally} by a polynomial of degree $p$ in a variable $j$, which
measures the distance between $y_{t}$ and its neighboring observations
$y_{t+j}, j=-m,\ldots,m$. This is equivalent to estimating the
trend-cycle $\hat{g}_{t}$ as a weighted moving average as follows:
%
%e3.2 #&#
\begin{equation}
\hat{g}_{t}=\sum_{j=-m}^{m}w_{j}y_{t+j}=
\mathbf{w}'\mathbf{y}, \qquad t=m+1, \ldots,N-m,
\end{equation}
where $\mathbf{w}'=[
w_{-m}\  \cdots\  w_{0}\  \cdots\  w_{m}]$ contains
the weights to be applied to the input data $\mathbf{y}'=[
y_{t-m}\  \cdots\  y_{t}\  \cdots\  y_{t+m}]$ to
get the estimate $\hat{g}_{t}$ for each point in time.

Several
nonparametric estimators, based on different sets of weights $\mathbf
{w}$, have been developed in the literature. The Henderson filter
[\citet{Hen16}, Kenny and Durbin (\citeyear{KenDur82}), \citet{Loa99,LadQue01}] results from fitting a cubic
polynomial to the input values $\mathbf{y}$ by means of weighted least
squares, that is,
%
%e3.3 #&#
\begin{equation}
\label{ls}\min_{\bolds\beta} [\mathbf {y}-\mathbf{X}\bolds\beta
]'\mathbf{W} [\mathbf {y}-\mathbf{X}\bolds\beta ],
\end{equation}
%
%$$
%\sum_{j=-m}^{m}W_{j}[y_{t+j}-a_{0}-a_{1}j-a_{2}j^{2}-a_{3}j^{3}]^{2}$$
where
\[
\mathbf{X}=\left[ \matrix{1&-m&m^2&-m^3
\vspace*{2pt}\cr
1&-(m-1)&(m-1)^2&-(m-1)^3
\vspace*{2pt}\cr
\vdots &\cdots&\cdots&\vdots
\vspace*{2pt}\cr
1&0&0&0
\vspace*{2pt}\cr
\vdots&\cdots&\cdots&\vdots
\vspace*{2pt}\cr
1&(m-1)&(m-1)^2&(m-1)^3
\vspace*{2pt}\cr
1&m&m^2&m^3}
 \right],\qquad
\bolds\beta=\left[ %
\matrix{\beta_{0}
\vspace*{2pt}\cr
\beta_{1}
\vspace*{2pt}\cr
\beta_{2}
\vspace*{2pt}\cr
\beta_{3} }
 \right],
\]
and $\mathbf{W}=\operatorname{diag}(W_{-m}, \ldots, W_{0}, \ldots, W_{m})$ with
generic element $W_{j} \propto\{(m+1)^{2}-j^{2}\}\{(m+2)^{2}-j^{2}\}\{
(m+3)^{2}-j^{2}\}$, chosen as to minimize the sum of squares of the
third differences of the weights $\mathbf{w}$. As discussed by \citet
{Loa99}, the latter are given by the product of a cubic polynomial
$\phi(j)$ and $W_{j}$, such that
\[
\hat{g}_{t}=\sum_{j=-m}^{m}
\phi(j)W_{j}y_{t+j}.
\]
For large $m$, \citet{Loa99} provided an equivalent kernel
representation of the weights by showing that $W_{j}$ can be
approximated by the triweight function $m^{6}(1-(j/m)^2)^{3}$, such
that the weight diagram is approximately
$(315/512)(3-11(j/m)^2)(1-(j/m)^2)^{3}$.

Different kernel
characterizations of the Henderson filter have been derived by
\citeauthor{DagBia08} (\citeyear
{DagBia08,DagBia13}) based on the Reproducing Kernel Hilbert Space
(RKHS) methodology. A RKHS is a Hilbert space characterized by a kernel
that reproduces, via an inner product, every function of the space. It
follows that a kernel estimator of order $p$ can always be decomposed
into the product of a reproducing kernel $R_{p-1}$, belonging to the
space of polynomials of degree at most $p-1$, and a probability density
function $f_{0}$ with finite moments up to order $2p$ [\citet{Ber93}].
In this context, the equivalent kernel representation of the Henderson
filter is given by
%
%e3.4 #&#
\begin{equation}
\label{reprker} K_{4}(t)=\sum_{i=0}^{3}P_{i}(t)P_{i}(0)f_{0}(t),\qquad
t \in[-1,1],
\end{equation}
where $f_{0}$ is the density function, defined on $[-1,1]$, obtained
through normalization of $W_{j}$, and the $P_{i}$ are the corresponding
orthonormal polynomials. Equivalently, the kernel in (\ref{reprker})
can be written as
%
%e3.5 #&#
\begin{equation}
\label{reprker2} K_{4}(t)= \frac{\det(\mathbf{H}_{4}^{0}[1, \mathbf{t}])}{\det(\mathbf
{H}_{4}^{0})}f_{0}(t),\qquad t
\in[-1,1],
\end{equation}
where $\mathbf{H}_{4}^{0}$ is the Hankel matrix whose elements are the
moments of $f_{0}$, that is, $\mu_{r}=\int_{-1}^{1}t^{r}f_{0}(t)\,\mathrm
{d}t$. In particular, the first row contains the moments from $\mu_{0}$
to $\mu_{3}$, whereas the last row those from $\mu_{3}$ to $\mu_{6}$.
$\mathbf{H}_{4}^{0}[1, \mathbf{t}]$ is the matrix obtained by replacing
the first column of $\mathbf{H}_{4}^{0}$ by the vector $\mathbf{t} = [
1\  t\   t^{2}\  t^{3}
]'$.

The density $f_{0}$ depends on $W_{j}$, hence, on the length of
the filter, and it needs to be determined any time that $m$ changes.
The kernel representation based on the triweight function allows to
overcome such limitation, but \citet{DagBia08} have found that the
biweight function $f_{0B}(t)=(15/16)(1-t^2)^2, t \in[-1,1]$, provides
a better approximation for Henderson filters of short length, say,
between 5 to 23 terms which are those used by statistical agencies
[see also \citet{BiaQue10}].

When applied to real data, the symmetric filter weights are derived as
follows:
%
%e3.6 #&#
\begin{equation}
\label{wei_0}w_{j}=\frac{K_{4}(j/b)}{\sum_{j=-m}^{m}K_{4}(j/b)},\qquad j=-m, \ldots, m,
\end{equation}
where $b$ is a time-invariant global bandwidth parameter (same for all
$t=m+1,\ldots,N-m$) selected to ensure a symmetric filter of length
$2m+1$. The bandwidth parameter relates the discrete domain of the
filter, that is, $\{-m,\ldots,m\}$, with the continuous domain of the
kernel function, that is, $[-1,1]$. The weights given in (\ref{wei_0})
can be also rewritten in matrix form as follows.
%Hence, in the following, we consider the kernel representation of the
%Henderson filter as given in eq. (\ref{reprker}), where $f_{0}$ is the
%biweight density function.

%pr3.1 #&#
\begin{prop}\label{454521521454} The weights $\mathbf{w}$ derived using the kernel function
in (\ref{reprker}) admit the following representation:
%
%e3.7 #&#
\begin{equation}
\label{wei}\mathbf{w}'=\mathbf{e}_{1}'{
\mathbf {H}_{s}}^{-1}\mathbf{X}_{b}'
\mathbf{F}_{b},
\end{equation}
where $\mathbf{e}_{1}'=[
1\  0\  0\  0
]$, $\mathbf{H}_{s}=\mathbf{H}_{4}^{0}[1, \mathbf{S}]$ with $\mathbf{S}'=[
S_{0}\  0\  S_{2}\  0
]$, being $S_{r}=  b^{-1}\sum_{j=-m}^{m}(j/b)^{r}f_{0}(j/b)$ the discrete
approximation of $\mu_{r}$, and $b$ the bandwidth parameter. In
addition, $\mathbf{X}_{b}$ has the same form as $\mathbf{X}$ in (\ref
{ls}), but with generic row given by $[
1\  j/b\  (j/b)^2\  (j/b)^3
]$, $j=-m, \ldots, m$, and $\mathbf{F}_{b}=\operatorname{diag}(1/b
f_{0B}(-m/b),\ldots, 1/bf_{0B}(m/b))$.
\end{prop}

 A formal proof of Proposition~\ref{454521521454} is provided in the \hyperref[app]{Appendix}.
It can be easily shown that the generic element of $\mathbf{w}$ is
given by
%
%e3.8 #&#
\begin{equation}
\label{sym}w_{j}= \biggl[\frac{\mu_{4}-\mu_{2} (
{j}/{b} )^{2}}{S_{0}\mu_{4}-S_{2}\mu_{2}} \biggr]
\frac
{1}{b}f_{0B} \biggl(\frac{j}{b} \biggr),\qquad j=-m,\ldots,
m.
\end{equation}
In this setting, once the length of the filter is selected, the choice
of the bandwidth parameter $b$ is fundamental. It has to be chosen to
ensure that only, say, $2m+1$ observations surrounding the target point
will receive nonzero weights as well as to approximate, as close as
possible, the continuous density function with the discrete one as well
as its moments. Indeed, we can separate (\ref{wei}) into two parts. One
concerns the discretization of the density function $f_{0}$ in terms of
adjacent rectangles, erected over discrete intervals, whose width is
determined by the bandwidth $b$. The second part corresponds to the
discretization of the reproducing kernel that depends on the discrete
moments $S_{0}$ and $S_{2}$. Of these two parts, the former plays the
most important role to approximate the continuous kernel given in (\ref
{reprker}) for the Henderson filter representation. Its bandwidth
parameter selection is done to guarantee specific inferential
properties of the trend-cycle estimators. In this regard,
\citeauthor{DagBia08} (\citeyear
{DagBia08,DagBia13}) used a time-invariant global bandwidth $b$ equal to
$m+1$, which gave excellent results.

%s3.1 #&#
\subsection{Asymmetric filters}\label{sec.1}
The derivation of the symmetric Henderson filter has assumed the
availability of $2m + 1$ input values centered at $t$. However, at the
end of the sample period, that is, $t=N-(m+1), \ldots, N$, only $2m,
\ldots, m+1$ observations are available, and asymmetric filters of the
same length have to be considered. Hence, at the boundary, the
effective domain of the kernel function $K_{4}$ is $[-1,q^{*}]$, with
$q^{*} \ll 1$, instead of $[-1,1]$ as for any interior point. This
implies that the symmetry of the kernel is lost, and it does not
integrate to unity on the asymmetric support
[$\int_{-1}^{q^{*}}K_{4}(t)\,\mathrm{d}t \neq1$]. Furthermore, the moment
conditions are no longer satisfied, that is, $\int_{-1}^{q^{*}}t^{i}K_{4}(t)\,\mathrm{d}t \neq0$, for $i=1,2,3$. To overcome
these limitations, several boundary kernels have been proposed in the
literature.

In the context of real time trend-cycle estimation, the
condition that the kernel function integrates to unity is essential,
whereas the unbiasedness property can only be satisfied with a great
increase in the variance of the estimates.
This is a consequence of the well-known trade-off between bias and
variance. This latter becomes very large because most of the
contribution to the real time trend-cycle estimates comes from the
current observation which gets the largest weight. Based on these
considerations,
\citeauthor{DagBia08} (\citeyear{DagBia08,DagBia13}) have suggested following the so-called ``cut
and normalize'' method [\citet{GasMul79,KyuSch98}], according to which
the boundary kernels $K_{4}^{q^{*}}$ are obtained by cutting the
symmetric kernel $K_{4}$ to omit that part of the function lying
between $q^{*}$ and 1, and by normalizing it on $[-1,q^{*}]$. That; that is,
%
%e3.9 #&#
\begin{equation}
\label{reprker3}K_{4}^{q^{*}}(t)=\frac{K_{4}(t)}{\int_{-1}^{q^{*}}K_{4}(t)\,\mathrm{d}t}=
\frac{\det(\mathbf{H}_{4}^{0}[1, \mathbf
{t}])f_{0B}(t)}{\det(\mathbf{H}_{4}^{0}[1, \bolds\mu^{q^{*}}])},\qquad t \in\bigl[-1,q^{*}\bigr],
\end{equation}
%
%\begin{flushright}$j=-m,\cdots,q; q=0,\cdots, m-1$\end{flushright}
where $\bolds\mu^{q^{*}}=[
\mu_{0}^{q^{*}}\  \mu_{1}^{q^{*}}\  \mu_{2}^{q^{*}}\  \mu_{3}^{q^{*}}
]$ with $\mu_{r}^{q^{*}}=\int_{-1}^{q^{*}}t^{r}f_{0B}(t)\,\mathrm{d}t$ being
proportional to the moments of the truncated biweight density $f_{0B}$
on the support $[-1,q^{*}]$, which from now on we simply refer to as
truncated moments.
%It can be noticed from eq. (\ref{reprker2}) that, for the last $m$
%points, $t=N,\cdots, N-m+1$, time-varying boundary kernels defined on
%different supports, ranging from -1 to $q^{*}=q/b_{q}$, $q=0, \cdots,
%m+1$, are considered.

 Applied to real data, the ``cut and normalize'' method yields
the following formula for the asymmetric weights:
%
%e3.10 #&#
\begin{equation}
\label{weasy} w_{q,j}=\frac{K_{4}^{q^{*}}(j/b_{q})}{\sum_{j=-m}^{q}K_{4}^{q^{*}}(j/b_{q})}=\frac{\det(\mathbf{H}_{4}^{0}[1,
\mathbf{j}/\mathbf{b}_{q}])(1/b_{q})f_{0B}(j/b_{q})}{\det(\mathbf{H}_{a})},
\end{equation}
for $j=-m,\ldots,q$, and $q=0,\ldots, m-1$, where $b_{q}, q=0, \ldots,
m-1$, is the local bandwidth, specific for each asymmetric filter. As
before, $b_{q}$ allows us to relate the discrete domain of the filter,
that is, $\{-m,\ldots,q\}$, for each $q= 0, \ldots, m-1$, to the
continuous domain of the kernel function, that is, $[-1,q^{*}]$.
Furthermore, $\mathbf{j}/\mathbf{b}_{q}=[
1\  (j/b_{q})\  (j/b_{q})^2\  (j/b_{q})^3
]$, and $\mathbf{H}_{a}=\mathbf{H}_{4}^{0} [1, \mathbf{S}^{q}
]$ with $\mathbf{S}^{q}=[
S_{0}^{q}\  S_{1}^{q}\  S_{2}^{q}\  S_{3}^{q}
]'$, and $S_{r}^{q}=\sum_{j=-m}^{q}(1/b_{q})(j/b_{q})^{r}f_{0B}(j/b_{q})$ the discrete
approximation of $\mu_{r}^{q^{*}}$.

%pr3.2 #&#
\begin{prop}\label{pro2} Each asymmetric filter $\mathbf
{w}_{q}=[w_{q,-m} \cdots w_{q,q}]'$ of length $(m+q+1)$, for $q=0,
\ldots, m-1$, admits the following matrix representation:
%
%e3.11 #&#
\begin{equation}
\label{wei2}\mathbf{w}_{q}'=\mathbf{e}_{1}'{
\mathbf {H}_{a}}^{-1}\mathbf{X}_{q}'
\mathbf{F}_{q}, \qquad q=0, \ldots, m-1,
\end{equation}
where $\mathbf{X}_{q}$ is a matrix of dimensions $(m+q+1) \times4$,
whose generic row is given by $\mathbf{j}/\mathbf{b}_{q}$, $j=-m, \ldots, q$,
and $\mathbf{F}_{q}=\operatorname{diag}((1/b_{q})f_{0B}(-m/b_{q}), \ldots,\break
(1/b_{q})f_{0B}(q/b_{q}))$. It can be easily shown that the generic
element of $\mathbf{w}_{q}$ is
%
%e3.12 #&#
\begin{equation}
\label{asym} w_{q,j}= \biggl[\frac{\mu_{4}-\mu_{2} ({j}/{b_{q}}
)^{2}}{S_{0}^{q}\mu_{4}-S_{2}^{q}\mu_{2}} \biggr]
\frac
{1}{b_{q}}f_{0B} \biggl(\frac{j}{b_{q}} \biggr),
\end{equation}
where $j=-m,\ldots, q$ and $q=0,\ldots, m-1$.
\end{prop}

 The proof of Proposition~\ref{pro2} is similar to that of
Proposition~\ref{454521521454} and, for space reasons, is omitted.

%s3.1.1 #&#
\subsubsection{Properties of the asymmetric filters}\label{sec3.1.1}
Since the trend-cycle estimates for the last $m$ data points do not use
$2m+1$ observations for any interior point, but $2m,2m-1, \ldots, m+1$
data, they are subject to revisions due to the following: (1)~new
observations entering in the estimation and (2) filter changes. As said
before, we will concentrate on the reduction of revisions due to filter
changes. The reduction of these revisions is an important property that
the asymmetric filters should possess together with a fast detection of
true turning points. In the specific case of the RKHS filters, (\ref
{asym}) shows how the asymmetric filter weights are related to the
symmetric ones given in (\ref{sym}). It is clear that the convergence
depends on the relationship between the two discretized biweight
density functions, truncated and nontruncated, jointly with the
relationship between their respective truncated $S_{r}^{q}$ and
untruncated $S_{r}$ discrete moments. The latter provide an
approximation of the continuous moments $\mu_{r}$, which improves as
the asymmetric filter length increases.
Similarly, the convergence of $S_{r}^{q}, q=0, \ldots, m$, to the
corresponding nontruncated moment $S_{r}$ depends on the length of the
asymmetric filter given by $q$ and on the local bandwidth $b_{q}$. It
should be noticed that $b_{q}$ plays a very important role in the
convergence property. For the last trend-cycle point weight, $ q=0$,
(\ref{asym}) reduces to
\[
w_{0,0} = \frac{\mu_{4}}{S_{0}^{0}\mu_{4}-S_{2}^{0}\mu_{2}}\frac{15}{16b_{0}}.
\]

\begin{figure}

\includegraphics{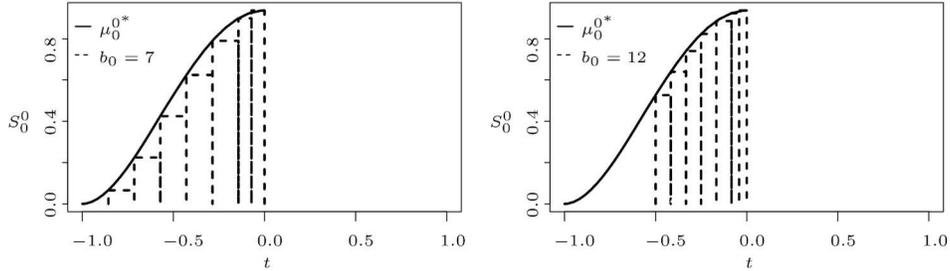}

\caption{Behavior of $S_{0}^{0}$ with $m=6$, $b_{0}=7$ (left) and
$b_{0}=12$ (right).}\label{Fig2}
\end{figure}

It is apparent that the larger $b_{0}$, the smaller is the weight given
to the last trend-cycle point. Since the sum of all the weights of the
last point asymmetric filter, $w_{0,-m}, \ldots, w_{0,0}$, must be
equal to one, this implies that the weights for the remaining points
are very close to one another. This can be seen in Figure~\ref{Fig2}
(right side) that shows, for $m=6$, the truncated continuous biweight
density function and its discretized version when $b_{0}$ is equal to
12. The opposite is observed when $b_{0}$ is smaller, as shown in the
same figure (left side) for $b_{0}$ equal to 7. Since a larger weight
is given to the last point, much smaller weights have to be assigned to
the remaining ones for all of them to add to one.
Next, we introduce time-varying local bandwidths to improve the
properties of the asymmetric filters in terms of size of revisions and
time delay to signal the upcoming of true turning points.

%f4 #&#

%s4 #&#
\section{Optimal bandwidth selection}\label{sec4}
The main effects induced by a linear filter on a given input are fully
described in the frequency domain by its transfer function
\[
\Gamma(\omega)=\sum_{j=-m}^{m}w_{j}
\exp(-i2\pi\omega j),\qquad \omega\in[-1/2,1/2],
\]
where, for better interpretation, the frequencies $\omega$ are given in
cycles for unit of time instead of radians. Here, $\Gamma(\omega)$
represents the Fourier transform of the filter weights, $w_{j}, j=-m,
\ldots, m$, and it relates the spectral density $h_{y}(\omega)$ and
$h_{g}(\omega)$ of the input and of the output, respectively, by
\[
h_{g}(\omega)=\Gamma(\omega)h_{y}(\omega).
\]
Thus, the transfer function $\Gamma(\omega)$ measures the effect of the
filter on the total variance of the input at different frequencies. It
is generally expressed in polar coordinates
%
%e4.1 #&#
\begin{equation}
\Gamma(\omega)=G(\omega) \exp\bigl(-i 2 \pi\phi(\omega)\bigr),
\end{equation}
so that the impact of the filter on a (complex-valued) series
$y_{t}=\exp(i 2 \pi\omega t)$, for $\omega\in[-1/2,1/2]$, is
\begin{eqnarray*}
\hat{g}_{t}&=&\Gamma(\omega)\exp(i 2 \pi \omega t)
\\
&=&G(\omega)\exp\bigl(-i 2\pi\phi(\omega)\bigr)\exp(i 2 \pi\omega t)
\\
&=&G(\omega)\exp\bigl\{i 2\pi\bigl[\omega t - \phi(\omega)\bigr]\bigr
\}.
\end{eqnarray*}
$G(\omega)=|\Gamma(\omega)|$ is the gain of the filter, which measures
the amplitude of the output for a sinusoidal input of unit amplitude,
whereas $\phi(\omega)$ is the phase function, which shows the shift in
phase of the output compared with the input.
%It is clear that the linear filter transforms a harmonic input series
%in two different ways: (a) multiplying it by an amplitude coefficient
%$G(\omega)$, and (b) shifting it in time by $\phi(\omega)/(2\pi
%\omega)$ with direct effects on its timeliness.
Hence, the transfer function plays a fundamental role to measure that
part of the total revisions due to filter changes.

The measure of total revisions introduced by \citet{Mus64} is
%
%e4.2 #&#
\begin{equation}
\label{revision}E \Biggl[\sum_{j=-m}^{q}w_{q,j}y_{t-j}-
\sum_{j=-m}^{m}w_{j}y_{t-j}
\Biggr]^{2},\qquad q=0, \ldots, m-1,
\end{equation}
where, in our case, $w_{q,j}$ and $w_{j}$ are given by (\ref{asym}) and
(\ref{sym}), respectively. This criterion can be expressed in the
frequency domain as follows: %\citet{Wil05,Wil08}
%
%e4.3 #&#
\begin{eqnarray}\label{mstr}
&&E \Biggl[\sum_{j=-m}^{q}w_{q,j}e^{i 2 \pi\omega
(t-j)}
-
\sum_{j=-m}^{m}w_{j}e^{i 2 \pi\omega(t-j)}
\Biggr]^{2}\nonumber\\
&&\qquad=E \bigl[\bigl(\Gamma_{q}(\omega)-\Gamma(\omega)
\bigr)e^{i 2 \pi\omega
t} \bigr]^{2}
\\
 &&\qquad=\int_{-1/2}^{1/2}\bigl|
\Gamma_{q}(\omega)-\Gamma(\omega)\bigr|^2 e^{i 4 \pi
\omega t}
h_{y}(\omega)\,\mathrm{d} \omega,\nonumber
\end{eqnarray}
where $h_{y}(\omega)$ is the unknown spectral density of $y_{t}$,
whereas $\Gamma_{q}(\omega)$ and $\Gamma(\omega)$ are the transfer
functions corresponding to the asymmetric and symmetric filters,
respectively. Similarly to (\ref{revision}), expression (\ref{mstr})
shows that, as new observations become available, revisions are due to
two sources: (a) the new innovations entering the input series, and (b)
changes in the asymmetric filters. In order to improve the current
trend-cycle prediction based on the asymmetric Henderson filters, we
study that part of the revisions due to asymmetric filter changes.
Because the estimation of the real time trend-cycle is done
concurrently, that is using all of the data up to and including the
most recent value, knowledge of the speed of convergence of the last
point trend-cycle filter to the central one gives valuable information
on how often the real time trend estimate should be revised.

The quantity $|\Gamma_{q}(\omega)-\Gamma(\omega)|^2$ accounts for the
revisions due to filter changes [\citeauthor{Dag82a} (\citeyear{Dag82a,Dag82b})], and it can be
decomposed using the law of cosines as follows:
%
%e4.4 #&#
\begin{eqnarray}\label{spec}
\bigl|\Gamma_{q}(\omega)-\Gamma(\omega)\bigr|^2
&=&\bigl|G_{q}(\omega)-G(\omega)\bigr|^2 + 2 G_{q}(
\omega)G(\omega) \bigl[1 - \cos \bigl(\phi_{q}(\omega)\bigr) \bigr]
\nonumber
\\[-8pt]
\\[-8pt]
\nonumber
&=& \bigl|G_{q}(\omega)-G(\omega)\bigr|^2 + 4
G_{q}(\omega)G(\omega) \sin \biggl(\phi_{q} \biggl(
\frac{\omega
}{2} \biggr) \biggr)^{2},
\end{eqnarray}
where the phase shift for the symmetric filter is equal to 0 or $\pm\pi
$, and where $1-\cos(\phi_{q}(\omega))=2 \sin (\phi_{q}(\omega
/2) )^2$. Based on (\ref{spec}), the mean square filter revision
error can be expressed as follows:
%
%
%e4.5 #&#
\begin{eqnarray}\label{msre}
2\int_{0}^{1/2}\bigl|\Gamma_{q}(
\omega)-\Gamma (\omega)\bigr|^2 \,\mathrm{d} \omega&=& 2\int
_{0}^{1/2}\bigl|G_{q}(\omega)-G(\omega
)\bigr|^2 \,\mathrm{d} \omega
\nonumber
\\[-8pt]
\\[-8pt]
\nonumber
&&{}+ 8\int_{0}^{1/2} G_{q}(
\omega )G(\omega) \sin \biggl(\phi \biggl(\frac{\omega}{2} \biggr)
\biggr)^{2}\, \mathrm {d} \omega.
\end{eqnarray}
The first component reflects the part of the total mean square filter
error which is attributed to the amplitude function of the asymmetric
filter. On the other hand, the second term measures the distinctive
contribution of the phase shift. The term
$G_{q}(\omega)G(\omega)$ is a scaling factor which accounts for the
fact that the phase function is dimensionless, that is, it does
not convey level information [\citet{Wil08}].

As previously discussed, once the length of the filter is chosen, the
properties of the asymmetric filters derived in RKHS are strongly
affected by the choice of the time-varying local bandwidths $b_{q},
q=0, \ldots,m-1$. Here, we propose several criteria for bandwidth
selection based on (\ref{msre}), and analyze the properties of the
corresponding optimal filters.
We define as optimal a filter that minimizes both revisions and time
delay to detect a true turning point. The LHS of (\ref{msre}) is a
measure of total filter revision that provides the best compromise
between the amplitude function of the asymmetric filter (gain) and its
phase function (time displacement) [\citeauthor{Dag82a} (\citeyear{Dag82a,Dag82b}), \citet{DagLan87}].
Optimal asymmetric filters in this sense can be derived using local
bandwidth parameters selected according to the following criterion:
%
%e4.6 #&#
\begin{equation}
\label{total}b_{q,\Gamma}= \min_{b_{q}} \sqrt{2\int
_{0}^{1/2}\bigl|\Gamma_{q}(\omega)-\Gamma(
\omega)\bigr|^2 \,\mathrm{d} \omega}.
\end{equation}
Based on the decomposition of the total filter revision error provided
in (\ref{msre}), further bandwidth selection criteria can be defined by
emphasizing more the gain or phase shift effects, and/or by attaching
varying importance to the different frequency components, depending on
whether they appear in the spectrum of the initial time series or not.
In the context of smoothing a monthly input, the frequency domain
$\Omega=\{0 \leq\omega\leq0.50\}$ can be partitioned in two main
intervals: (1)~$\Omega_{S}=\{0 \leq\omega\leq0.06\}$ associated with
cycles of 16 months or longer attributed to the signal (trend-cycle) of
the series, and (2) $\bar{\Omega}_{S}=\{0.06 < \omega\leq0.50\}$
corresponding to short cyclical fluctuations attributed to the noise.

We derive a class of optimal asymmetric filters based on bandwidth
parameters $b_{q}, q=0, \ldots, m-1$, selected as follows:
%
%e4.7 #&#
\begin{equation}
\label{gain} b_{q,G}= \min_{b_{q}} \sqrt{2
\int_{0}^{1/2}\bigl|G_{q}(\omega)-G(
\omega)\bigr|^2 \,\mathrm{d} \omega}
\end{equation}
and
%
%e4.8 #&#
\begin{equation}
\label{phi2} b_{q,\phi}= \min_{b_{q}} \sqrt{ 2 \int
_{\Omega_{S}} G_{q}(\omega)G(\omega) \bigl[1 - \cos\bigl(
\phi_{q}(\omega )\bigr) \bigr]\,\mathrm{d} \omega}.
\end{equation}
It has to be noticed that the minimization of the phase error in (\ref
{phi2}) is very close to minimizing the average phase shift in month
for the signal, that is,
%
%e4.9 #&#
\begin{equation}
\label{phi}b_{q,\phi}= \min_{b_{q}} \biggl[
\frac
{1}{0.06} \int_{\Omega_{S}} \frac{\phi(\omega)}{2 \pi\omega} \,\mathrm {d}
\omega \biggr].
\end{equation}
%
%Therefore, criteria (\ref{phi2}) and (\ref{phi}) provides the same set
%of asymmetric filters.
Table~\ref{band} illustrates the bandwidth parameters $b_{q,\Gamma}$,
$b_{q,G}, b_{q,\phi}, q=0,\ldots,m-1$, derived as minimizers of (\ref
{total}), (\ref{gain}) and (\ref{phi}), respectively, corresponding to
the 9-, 13- and 23-term symmetric filters. %\begin{center}

%t1 #&#
\begin{table}
\tabcolsep=0pt
\caption{Optimal bandwidth values selected for each of the biweight
asymmetric filters corresponding to the 9-, 13- and 23-term Henderson
symmetric filters}\label{band}
\begin{tabular*}{\textwidth}{@{\extracolsep{\fill}}ld{2.2}d{2.2}d{2.2}d{2.2}d{2.2}d{2.2}d{2.2}d{2.2}d{2.2}d{2.2}d{2.2}@{}}
\hline
$\bolds{q}$ &\textbf{0} &\textbf{1} &\textbf{2} & \textbf{3} & &&&&&& \\
\hline
$b_{q, \Gamma}$ & 6.47 &5.21 &4.90 &4.92 & & &&&& & \\
$b_{q,G}$ & 8.00 &5.67 & 4.87 & 4.90 & & &&&& & \\
$b_{q,\phi}$ & 4.01 &4.45 & 5.97 & 6.93 & & &&&& & \\[3pt]
\hline
$\bolds{q}$ &\textbf{0} &\textbf{1} &\textbf{2} & \textbf{3} & \textbf{4} & \textbf{5} &&&& &\\
\hline
$b_{q, \Gamma}$& 9.54 &7.88 &7.07 &6.88 &6.87&6.94&&&&&\\
$b_{q,G}$& 11.78 &9.24 & 7.34 & 6.85 & 6.84 & 6.95 &&&&& \\
$b_{q,\phi}$& 6.01 &6.01 & 7.12 & 8.44 & 9.46 & 10.39 &&&&& \\[3pt]
\hline
$\bolds{q}$ &\textbf{0} &\textbf{1} &\textbf{2} & \textbf{3} & \textbf{4} & \textbf{5} &\textbf{6}&\textbf{7}&\textbf{8}&\textbf{9}&\textbf{10} \\
\hline
$b_{q, \Gamma}$&
17.32&15.35&13.53&12.47&12.05&11.86&11.77&11.77&11.82&11.91&11.98\\
$b_{q,G}$&21.18&18.40&16.07&13.89&12.44&11.90&11.72&11.73&11.83&11.92&11.98\\
$b_{q,\phi
}$&11.01&11.01&11.01&11.01&11.41&13.85&15.13&16.21&17.21&18.15&19.05\\
\hline
\end{tabular*}
\end{table}

\begin{figure}

\includegraphics{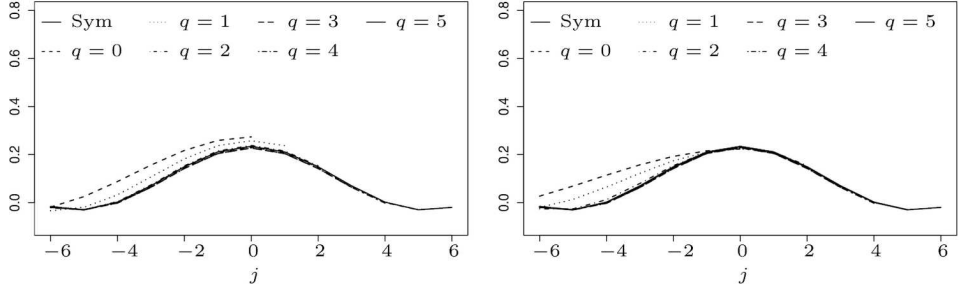}

\caption{Time path of the asymmetric filters based on $b_{q, \Gamma}$
(left), $b_{q,G}$ (right) corresponding to the 13-term symmetric
filter.}\label{Fig3}
\end{figure}

 It can be noticed that, as $q$ approaches $m$, the bandwidth
parameters selected to optimize the criteria (\ref{total}) and (\ref
{gain}) get closer to $m+1$, that is the global bandwidth considered
for the symmetric Henderson filter. Hence, based on the relationships
between truncated and untruncated discrete biweight density functions
and respective discrete moments previously discussed, the asymmetric
filters based on $b_{q, \Gamma}$ and $b_{q,G}$, $q=0, \ldots, m-1$,
should be characterized by a fast convergence to the symmetric filter.
This is confirmed by Figure~\ref{Fig3} that illustrates, as an example,
the time path of these filters corresponding to the 13-term symmetric
one. Other filter lengths have been considered, but, for space reasons,
we only show the results for the 13-term filter. However, similar
conclusions can be drawn for different filter lengths.

%f5 #&#

%f6 #&#
\begin{figure}[b]

\includegraphics{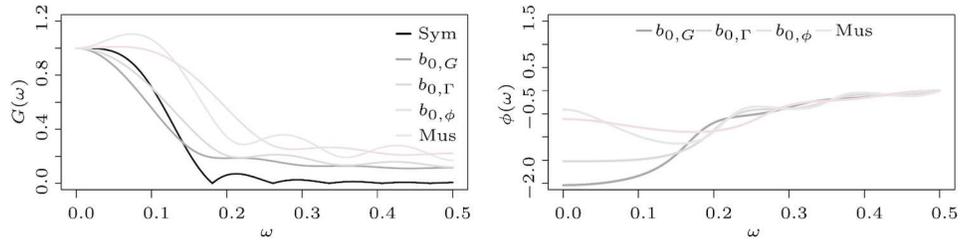}

\caption{Gain (left) and phase shift (right) functions for the last
point asymmetric filters based on $b_{0, \Gamma}$, $b_{0,G}$ and $b_{0,
\phi}$ compared with the last point Musgrave filter.}\label{Fig5}
\end{figure}

%f7 #&#
\begin{figure}

\includegraphics{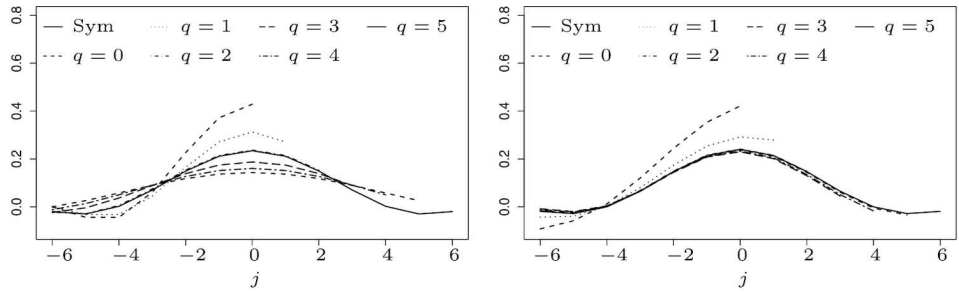}

\caption{Time path of the asymmetric filters based on $b_{q, \phi}$
(left) and of the Musgrave asymmetric filters (right) corresponding to
the 13-term symmetric filter.}\label{Fig4}
\end{figure}

The asymmetric filters based on $b_{q, \Gamma}$ and $b_{q,G}$, $q=0,
\ldots, m-1$, converge very fast to the symmetric filter, particularly
after the previous to the last point, with the main differences
observed for the last point filters. For these latter, the different
behavior is analyzed in the frequency domain in Figure~\ref{Fig5}, that
shows the corresponding gain and phase shift functions. It can be
noticed that, as expected, the filter whose bandwidth $b_{0,G}$ is
derived as minimizer of (\ref{gain}) shows a gain function closer to
that of the symmetric Henderson filter than the one based on
$b_{0,\Gamma}$, suppressing more noise at the highest frequencies, and
it reproduces very well the signal in the lower frequency band.

In
terms of phase shift or time delay, the filters that behave better are
the ones based on the bandwidth parameters selected to minimize the
average phase shift in months over the signal domain. However, as shown
in Figure~\ref{Fig4}, their time path is only very close to that of the
filters derived by \citet{Mus64} up to $q=2$, but there is no monotonic
convergence of these asymmetric filters to their final one. This
property is reflected in their phase shift function that, for the last
point filter, is illustrated in Figure~\ref{Fig5}. As already said, the
Musgrave filters are based on the minimization of the mean squared
revision between the final estimates, obtained by the application of
the symmetric filter, and the preliminary estimates, obtained by the
application of an asymmetric filter, subject to the constraint that the
sum of the weights is equal to one [\citet{Lan85,Doh01}]. These filters
have the good property of fast detection of turning points.
%\begin{figure}[!htpb]
%\caption{Gain (left) and phase shift (right) functions for the
%previous to the last point asymmetric filters based on $b_{1,
%\Gamma}$, $b_{1,G}$, and $b_{1, \phi}$ compared with the previous to
%the last point Musgrave filter.}\label{Fig6} %
%
%
%\psfrag{GF}{\tiny$G(\omega)$}\psfrag{PF}{\tiny$\phi(\omega)$}
%\psfrag{w}{\tiny$\omega$}\psfrag{S}{\tiny Sym}\psfrag{G}{\tiny
%$b_{0,G}$}\psfrag{T}{\tiny$b_{0,\Gamma}$}\psfrag{P}{\tiny$b_{0,
%\phi}$}\psfrag{F}{\tiny$b=7$}\psfrag{M}{\tiny Mus}
%\psfrag{0.0}{\tiny$0.0$}\psfrag{0.1}{\tiny$0.1$}\psfrag{0.2}{
%\tiny$0.2$}\psfrag{0.3}{\tiny$0.3$}\psfrag{0.4}{\tiny$0.4$}
%\psfrag{0.5}{\tiny$0.5$}\psfrag{0.8}{\tiny$0.8$}\psfrag{1.2}{
%\tiny$1.2$}
%\psfrag{-2.0}{\tiny$-2.0$}\psfrag{-0.5}{\tiny$-0.5$}\psfrag{1.5}{
%\tiny$1.5$}
%\includegraphics[width=6cm,height=4cm]{gainPL.eps}
%\includegraphics[width=6cm,height=4cm]{phasePL.eps}
%\end{figure}

As we can see, both the last point Musgrave filter and the one based on
$b_{0,\phi}$ produce almost one half of the phase shift introduced by
the filter based on $b_{0,\Gamma}$ and a quarter of the one introduced
by the filter based on $b_{0,G}$ at the signal frequency band. However,
the reduced phase shift produced by these two filters is compensated by
larger revisions introduced in the final estimates. Indeed, as shown by
the corresponding gain functions, the last point Musgrave filter and
the one based on $b_{0,\phi}$ suppress much less noise than the filters
obtained through minimization of (\ref{total}) and (\ref{gain}).
Furthermore, the Musgrave filter has the worst performance since it
introduces a large amplification of the power attributed to the trend
and suppresses less noise.\looseness=1

%s5 #&#
\section{Application to the US economy}\label{sec5}
%The asymmetric filters previously derived can be applied in many
%fields, such as, economics, finance, health, hydrology, meteorology,
%criminology, physics, labor markets, utilities, and so on. In fact, in
%any time series where the impact of trend plus cyclical variations is
%of relevance.
We have chosen a set of leading, coincident and lagging indicators of
the US economy to illustrate some of the potential gains of using these
new asymmetric filters.
Time series that exhibit a turning point before the economy as a whole
are called leading indicators, whereas those that change direction
approximately at the same time are called coincident indicators. The
lagging indicators are those that usually change direction after the
whole economy does. The composite indexes are typically reported in
financial and trade media. The series analyzed in this study are
obtained from the St. Louis Federal Reserve Bank database, the Bureau
of Labor Statistics, the Conference Board and the National Bureau of
Economic Research (NBER). They are all final vintages data in the sense
that they will no longer be revised.
We have chosen the following as leading indicators:
\begin{itemize}[--]
\item[--] Composite index
of ten leading indicators ($2010=100$).\footnote{The index is rebased to
average 100 in 2010. The history of the index is multiplied by 100 and
divided by the average for the twelve months of the based year,
currently 2010.}

\item[--] Average weekly overtime hours, manufacturing.

\item[--] New
orders for durable goods.

\item[--] New orders for nondefense capital goods.

\item[--]
New private housing units authorized by building permits.

\item[--] Stock
prices, S\&P common stocks.

\item[--] Money supply, M2.

\item[--] Interest rate spread,
10-year treasury bonds less federal funds.

\item[--] Index of consumer
expectation (University of Michigan).
\end{itemize}

We consider the following as
coincident indicators:
\begin{itemize}[--]
\item[--] Composite index of four coincident indicators
($2010=100$).

 \item[--] Employees on nonagricultural payrolls.

 \item[--] Personal income
less transfer payments.

\item[--] Industrial production index.

\item[--] Manufacturing
and trade sales.
\end{itemize}
Finally, the lagging indicators treated are as
follows:
\begin{itemize}[--]
\item[--] Composite index of seven lagging indicators ($2010=100$).

\item[--] Average duration of unemployment, weeks.

\item[--] Ratio, manufacturing and
trade inventory to sale.

\item[--] Change in labor cost per unit of output,
manufacturing.

\item[--] Commercial and industrial loans outstanding.
\end{itemize}
The asymmetric filters derived following the RKHS methodology versus
the Musgrave filters, applied in conjunction with the symmetric
Henderson filter, are evaluated as follows.

%s5.1 #&#
\subsection{Reduction of revision size in real time short-term trend estimates}\label{sec5.1}
The reduction of revisions in real time trend-cycle estimates is very
important because the estimates are preliminary and often used to
assess the current stage of the economy. Statistical agencies and major
users of these indicators are reluctant to large revisions because
these can lead to wrong decision taking and policy making concerning
the current economic situation. The series considered are all
seasonally adjusted, where also trading day variations and extreme
values have been removed if present. The indicators are series of
different length, but the periods selected sufficiently cover the
various lengths published for these series. For each series, the length
of the filters is selected according to the $I/C$ (noise to signal)
ratio, as classically done in the X11/X12ARIMA procedure [\citet
{LadQue01}]. In the sample, the ratio ranges from 0.20 to 1.98, hence
filters of length 9 and 13 terms are applied.

%t2 #&#
\begin{table}
\tabcolsep=0pt
\caption{Ratio of the mean square percentage revision errors of the
last point asymmetric filters based on $b_{0,G}$, $b_{0,\Gamma}$ and
$b_{0,\phi}$, and the last point Musgrave filter}\label{Tab3}
\begin{tabular*}{\textwidth}{@{\extracolsep{\fill}}lcccc@{}}
\hline
\multicolumn{1}{@{}l}{\textbf{Macro-}}\\
\multicolumn{1}{@{}l}{\textbf{area}}& \multicolumn{1}{c}{\textbf{Series}} & \multicolumn{1}{c}{$\bolds{\frac{b_{0,G}}{\mathbf{Mus}}}$}&\multicolumn{1}{c}{$\bolds{\frac{b_{0,\Gamma}}{\mathbf{Mus}}}$} &\multicolumn{1}{c@{}}{$\bolds{\frac{b_{0,\phi}}{\mathbf{Mus}}}$}\\
\hline\\
\multicolumn{1}{@{}l}{Leading} &\multicolumn{1}{l}{Composite index of ten
leading indicators }& 0.503 & 0.643&0.933\\
&\multicolumn{1}{l}{Average weekly overtime hours: Manufacturing}&
0.492& 0.630&0.922\\
&\multicolumn{1}{l}{New orders for durable goods}&0.493&0.633&0.931\\
&\multicolumn{1}{l}{New orders for nondefense capital
goods}&0.493&0.633&0.931\\
&\multicolumn{1}{l}{New private housing units authorized by building
permits}&0.475 &0.651&0.927\\
&\multicolumn{1}{l}{S\&P 500 stock price index}&0.454&0.591&0.856 \\
&\multicolumn{1}{l}{M2 money stock}&0.508&0.655&0.932\\
&\multicolumn{1}{l}{10-year treasury constant maturity
rate}&0.446&0.582&0.849\\
&\multicolumn{1}{l}{University of Michigan: Consumer sentiment}&0.480&0.621&0.912\\[3pt]
\multicolumn{1}{@{}l}{Coinci-} &\multicolumn{1}{l}{Composite index of four coincident indicators}&0.504&0.651&0.931\\
\multicolumn{1}{@{}l}{\quad dent} &\multicolumn{1}{l}{All employees: total nonfarm}&0.517&0.666 &0.951\\
&\multicolumn{1}{l}{Real personal income excluding current transfer
receipts}&0.484&0.627&0.903\\
&\multicolumn{1}{l}{Industrial production index}&0.477&0.616&0.884\\
&\multicolumn{1}{l}{Manufacturing and trade sales}&0.471&0.606&0.869\\[3pt]
\multicolumn{1}{@{}l}{Lagging} &\multicolumn{1}{l}{Composite index of
seven lagging indicators}&0.523&0.653&0.966\\
&\multicolumn{1}{l}{Average (mean) duration of
unemployment}&0.509&0.649&0.937\\
&\multicolumn{1}{l}{Inventory to sales ratio}&0.483&0.618&0.894\\
&\multicolumn{1}{l}{Index of total labor cost per unit of
output}&0.515&0.663&0.983\\
&\multicolumn{1}{l}{Commercial and industrial loans at all commercial
banks}&0.473&0.610&0.871\\
\hline
\end{tabular*}
\end{table}

The comparisons are
based on the relative filter revisions between the final symmetric
filter $S$ and the last point asymmetric filter $A$, that is,
%
%e5.1 #&#
\begin{equation}
R_{t}=\frac{S_{t}-A_{t}}{S_{t}},\qquad t=1,\ldots,N.
\end{equation}
For each series and for each estimator, we calculate the ratio between
the Mean Square Percentage Error (MSPE) of the revisions corresponding
to the filters derived following the RKHS methodology and those
corresponding to the last point Musgrave filter. For all the
estimators, the results illustrated in Table~\ref{Tab3} show that the
ratio is always smaller than one, indicating that the kernel last point
predictors, based on time-varying bandwidth parameters, introduce
smaller revisions than the Musgrave filter. This implies that the
estimates obtained by the former will be more accurate than those
derived by the application of the latter. In particular, as expected,
the best performance is shown by the filter based on the optimal
bandwidth $b_{0,G}$ derived to minimize the criterion (\ref{gain}). In
almost all the series its ratio with the last point Musgrave filter is
less than one half and, on average, around 0.489. This implies that
when applied to real data, the filter based on $b_{0,G}$ produces a
reduction of almost fifty percent of the revisions introduced in the
real time trend-cycle estimates given by the Musgrave filter. The
filter based on $b_{0,\Gamma}$, derived to minimize the size of total
filter revisions as defined by (\ref{total}), also performs very well
with more than thirty percent of revision reduction with respect to the
Musgrave filter. In this case, the ratio is greater than the one
corresponding to the filter based on $b_{0,G}$, but always less than
0.7 for all the series, being, on average, around 0.631. The filter
whose bandwidth parameter is selected to minimize the average phase
shift over the signal domain performs more similarly to the last point
Musgrave filter but still shows revisions reduction, on average, around
ten percent.

%s5.2 #&#
\subsection{Turning point detection}\label{sec5.2}

It is important that the reduction of revisions in real time
trend-cycle estimates is not achieved at the expense of increasing the
time lag to detect the upcoming of a true turning point. A turning
point is generally defined to occur at time $t$ if (\emph{downturn})
\[
y_{t-k} \leq\cdots\leq y_{t-1} > y_{t} \geq
y_{t+1} \geq\cdots\geq y_{t+m}
\]
or (\emph{upturn})
\[
y_{t-k} \geq\cdots\geq y_{t-1} < y_{t} \leq
y_{t+1} \leq\cdots\leq y_{t+m}.
\]

Following \citet{Zel91}, we have chosen $k=3$ and $m=1$ given the
smoothness of the trend-cycle data. For each estimator, the time lag to
detect the true turning point is
affected by the convergence path of its asymmetric filters $\mathbf
{w}_{q}, q=0, \ldots, m-1$, to the symmetric one $\mathbf{w}$.

To determine the time lag needed by an indicator to detect a true
turning point, we calculate the number of months it takes for the real
time trend-cycle estimate to signal a turning point in the same
position as in the final trend-cycle series. For the series analyzed in
this paper, the time delays for each estimator are shown in Table~\ref{Tab6}. It can be noticed that the filters based on the bandwidth
$b_{q,\phi}$ take two months (on average) as the Musgrave filters to
detect the turning point. This is due to the fact that, even if
$b_{q,\phi}$ filters are designed to be optimal in timeliness, their
convergence path to the symmetric filter is slower and not monotone.

On the other hand, the filters based on $b_{q,\Gamma}, q=0, \ldots,
m-1$, and $b_{q,G}, q=0, \ldots, m-1$, perform strongly better. In
particular, whereas the former detect the turning point with an average
time delay of 1.44 months, the latter takes 1.22 months.

The faster the upcoming of a turning point is detected, the faster new
policies can be applied to counteract the impact of the business cycle
stage. Failure to recognize the downturn in the cycle or taking a long
time delay to detect it may lead to the adoption of policies to curb
expansion when, in fact, a recession is already underway.

%t3 #&#
\begin{table}
\tabcolsep=0pt
\caption{Time lag in detecting true turning points for the asymmetric
filters based on $b_{q,G}$, $b_{q,\Gamma}$ and $b_{q,\phi}$, and the
Musgrave filters}\label{Tab6}
\begin{tabular*}{\textwidth}{@{\extracolsep{\fill}}lccccc@{}}
\hline
& & & & &\multicolumn{1}{l@{}}{\textbf{Mus-}}\\
\multicolumn{1}{@{}}{\textbf{Macro-area}}& \multicolumn{1}{c}{\textbf{Series}} &\multicolumn{1}{c}{$\bolds{b_{q,G}}$}&\multicolumn{1}{c}{$\bolds{b_{q,\Gamma}}$} &\multicolumn{1}{c}{$\bolds{b_{q,\phi}}$}&\multicolumn{1}{l@{}}{\textbf{grave}}\\
\hline
\multicolumn{1}{@{}l}{Leading} &\multicolumn{1}{l}{Composite index of ten leading indicators }& 1& 1&3&3\\
&\multicolumn{1}{l}{Average weekly overtime hours: Manufacturing}& 1&1&1&1 \\
&\multicolumn{1}{l}{New orders for durable goods}& 1& 2&3&2 \\
&\multicolumn{1}{l}{New orders for nondefense capital goods}& 1&
2&2&3\\
&\multicolumn{1}{l}{New private housing units authorized by building
permits}&2&2&3&3\\
&\multicolumn{1}{l}{S\&P 500 stock price index}&1&2&2&2\\
% &\multicolumn{1}{l}{M2 money stock}&0.722&0.819&0.977\\
&\multicolumn{1}{l}{10-year treasury constant maturity rate}& 1&1&1&2
\\
&\multicolumn{1}{l}{University of Michigan: Consumer sentiment}&1&
1&1&1\\[3pt]
\multicolumn{1}{@{}l}{Coincident}&\multicolumn{1}{l}{Composite index of
four coincident indicators}& 1& 1&2&2\\
&\multicolumn{1}{l}{All employees: total nonfarm}& 1& 1&1&2\\
&\multicolumn{1}{l}{Real personal income excluding current transfer
receipts}& 1& 1&1&1\\
&\multicolumn{1}{l}{Industrial production index}& 1& 1&1&1\\
&\multicolumn{1}{l}{Manufacturing and trade sales}&1&2&3&3\\[3pt]
\multicolumn{1}{@{}l}{Lagging} &\multicolumn{1}{l}{Composite index of
seven lagging indicators}&1& 1&3&3\\
&\multicolumn{1}{l}{Average (mean) duration of unemployment}&3&3&4&3 \\
&\multicolumn{1}{l}{Inventory to sales ratio}&1&1&1&2\\
&\multicolumn{1}{l}{Index of total labor cost per unit of
output}&2&2&3&2\\
&\multicolumn{1}{l}{Commercial and industrial loans at all commercial
banks}&1&1&1&1\\[6pt]
&Average time lag in months &1.22 &1.44&2.00&2.06\\
\hline
\end{tabular*}
\end{table}

\begin{figure}

\includegraphics{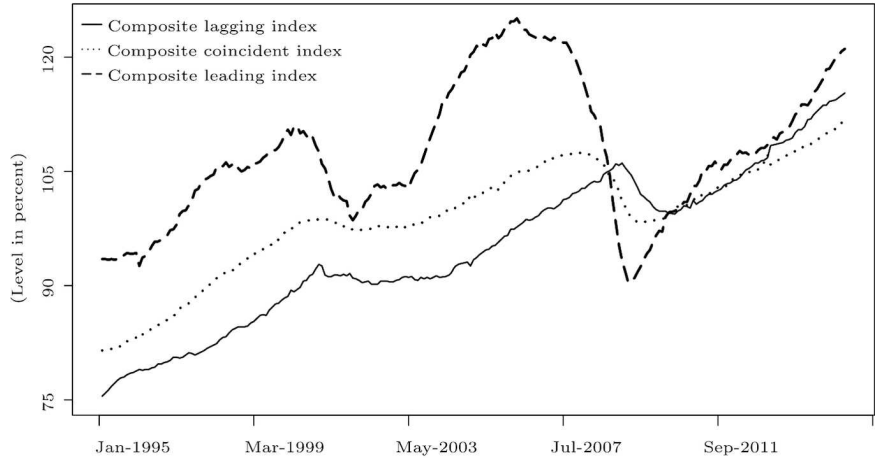}

\caption{Composite leading, coincident and lagging indicators of the
US economy ($2010=100$).}\label{Composite}
\end{figure}

To better highlight how the proposed filters perform when applied to
series that are impacted differently by the short-term trend, we look
at the revision path of the corresponding estimates. In this regard, we
compare the performance of the filters on the three composite
indicators, namely, leading, coincident and lagging, illustrated in
Figure~\ref{Composite} for the period January 1995--December 2014.
The composite index of ten leading indicators presents a deep turning
point on May 2009, whereas shallow turning points are shown by the
coincident and lagging composite indicators on August 2009 and May
2010, respectively.

\begin{figure} [b]

\includegraphics{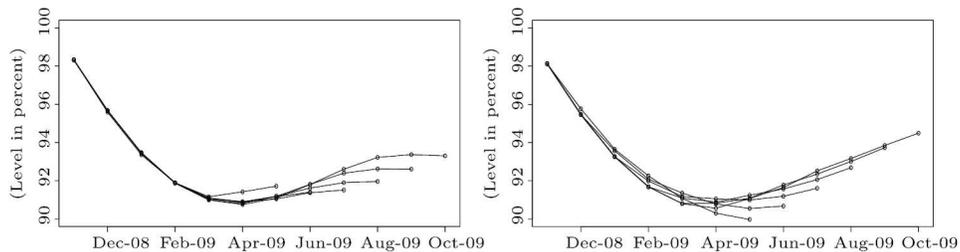}

\caption{Composite index of ten leading indicators ($2010=100$): revision
path of the May 2009 (turning point) estimate as one observation is
added at a time up to November 2009 (final estimate) using the
asymmetric kernels based on $b_{q,G}$ (left) and Musgrave (right)
filters, respectively.}\label{porclea}
\end{figure}

Figure~\ref{porclea} exhibits the behavior of the Musgrave filters
(right) and of the kernel filters based on $b_{q,G}$ in detecting the
May 2009 turning point of the composite leading index. In particular,
the figure shows the revision path of the last available point (May
2009) as we keep adding one observation at a time up to October 2009,
when the final estimate is achieved.
%

%f8 #&#

%f9 #&#

It can be noticed that after adding one month at the series ending at
May 2009, the turning point is clearly detected by the kernel filters,
whereas three months are required by the Musgrave ones.

\begin{figure}

\includegraphics{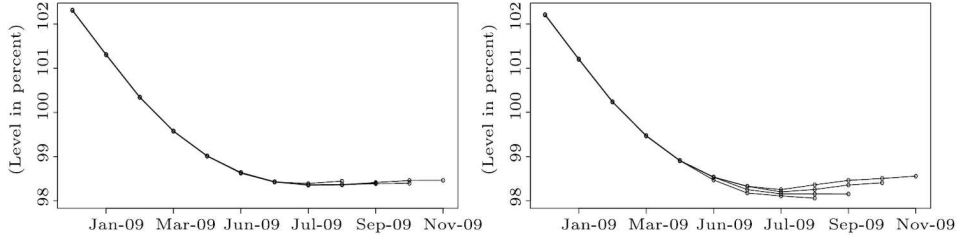}

\caption{Composite index of four coincident indicators ($2010=100$):
revision path of the August 2009 (turning point) estimate as one
observation is added at a time up to November 2009 (final estimate)
using the asymmetric kernels based on $b_{q,G}$ (left) and Musgrave
(right) filters, respectively.}\label{porcoin}
\end{figure}

\begin{figure}[b]

\includegraphics{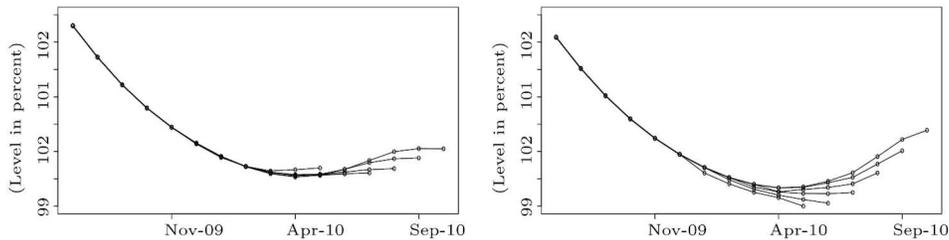}

\caption{Composite index of seven lagging indicators ($2010=100$):
revision path of the May 2010 (turning point) estimate as one
observation is added at a time up to November 2010 (final estimate)
using the asymmetric kernels based on $b_{q,G}$ (left) and Musgrave
(right) filters, respectively.}\label{porcmus}
\end{figure}

A similar pattern is observed in Figures~\ref{porcoin} and \ref
{porcmus} that are the ``porcupine'' graphs for the August 2009 and May
2010 turning points of the coincident and lagging composite indicators,
respectively. For both series, the kernel filters detect the turning
points after one month they have occurred, whereas the Musgrave filters
take two months for the former, and three months for the latter.
Hence, based on our previous considerations, the filters based on local
bandwidth parameters selected to minimize criterion (\ref{gain}) are
optimal, since they drastically reduce the total revisions by one half
with respect to the Musgrave filters and, similarly, almost by one half
the number of months needed to detect a true turning point.

%
%f10 #&#

%f11 #&#

%s6 #&#
\section{Discussion}\label{sec6}
This paper deals with the problem of assessing, in real time, the
direction of the short-term trend with an application to some key
indicators of the US economy. The linear asymmetric filters here
proposed are developed using the RKHS methodology. Given the length of
the RKHS asymmetric filter, its properties strongly depend on the
bandwidth parameter of the asymmetric kernel function from which the
filter weights are derived. Since the $m$ asymmetric filters
corresponding to a $2m+1$ symmetric filter are time varying, one for
each specific point, we are here proposing local time-varying bandwidth
parameters. We consider three main criteria for bandwidth selection in
order to determine an optimal smoother. An optimal filter is defined as
the one that minimizes revisions and time lag to detect the upcoming of
a true turning point.
The three main criteria of bandwidth parameter selection are
minimization of the following: (1) the distance between the gain
functions of asymmetric and symmetric filters, (2) the distance between
the transfer functions of asymmetric and symmetric filters, and (3) the
phase shift function over the domain of the signal.

We show theoretically that any of the three criteria produces
asymmetric trend-cycle filters to be preferred to those developed by
Musgrave concerning both size of revisions and time delay to detect the
upcoming of true turning points. To highlight how the proposed filters
perform when applied to series that are impacted differently by the
long-term trend, we look at the revision path of the corresponding
estimates. In this regard, we compare the performance of the filters on
three composite indicators, namely, leading, coincident and lagging.
The composite index of ten leading indicators presents a deep turning
point on May 2009, whereas shallow turning points are shown by the
coincident and lagging composite indicators on August 2009 and May
2010, respectively.
The real time trend-cycle filter calculated with the bandwidth
parameter that minimizes the distance between the asymmetric and
symmetric filters gain functions is to be preferred. This last point
trend-cycle filter reduces around one half the size of the total
revisions as well as the time delay to detect a true turning point with
respect to the Musgrave filter. The new set of asymmetric kernel
filters can be applied in many fields, such as economics, finance,
health, hydrology, meteorology, criminology, physics, labor markets,
utilities and so on, in fact, in any time series where the impact of
trend plus cyclical variations is of relevance. For interested readers,
the weight systems of these filters are given in the supplementary
material [\citet{supp}] for 9- and 13-term symmetric filters.

\begin{appendix}\label{app}
\section*{Appendix: Proof of Proposition~\texorpdfstring{\protect\ref{454521521454}}{3.1}}

As shown by Dagum and Bianconcini [(\citeyear{DagBia08}) and (\citeyear{DagBia13})], the symmetric filter
weights are derived as follows:
\[
w_{j}=\frac{K_{4}(j/b)}{\sum_{j=-m}^{m}K_{4}(j/b)},\qquad j=-m, \ldots, m,
\]
where $b$ is the time-invariant global bandwidth parameter (same for
all $t=m+1,\ldots,N-m$) selected to ensure a symmetric filter of length
$2m+1$. Based on (\ref{reprker2}), we obtain that
\begin{eqnarray*}
w_{j}&=&\frac{\det(\mathbf{H}_{4}^{0}[1,
\mathbf{j}/\mathbf{b}])({1}/{b})f_{0B} ({j}/{b} )}
{\sum_{j=-m}^{m}\det(\mathbf{H}_{4}^{0}[1, \mathbf{j}/\mathbf{b}])
({1}/{b})f_{0B} ({j}/{b} )}\\
&=&
\frac{\det(\mathbf{H}_{4}^{0}[1,
\mathbf{j}/\mathbf{b}])({1}/{b})f_{0B} ({j}/{b} )}{\det(\mathbf
{H}_{4}^{0}[1,\sum_{j=-m}^{m}\mathbf{j}/\mathbf{b}({1}/{b})f_{0B} (
{j}/{b} )])}
\\
&=&\frac{\det(\mathbf{H}_{4}^{0}[1, \mathbf
{j}/\mathbf{b}])({1}/{b})f_{0B} ({j}/{b} )}{\det(\mathbf
{H}_{4}^{0}[1,\mathbf{S}])}=\frac{\det(\mathbf{H}_{4}^{0}[1, \mathbf
{j}/\mathbf{b}])({1}/{b})f_{0B} ({j}/{b} )}{\det(\mathbf{H}_{s})},
\end{eqnarray*}
where $\mathbf{H}_{s}=\mathbf{H}_{4}^{0}[1,\mathbf{S}]$, with $\mathbf
{S}=[S_{0}^{b},0,S_{2}^{b},0]'$, and $S_{r}^{b}=0$ for odd $r$. The
expression above is exactly the same as we would obtain by solving for
$\hat{\beta}_{0}=\hat{g}_{t}$ the system of linear equations
\[
\mathbf{H}_{s} \bolds\beta=\mathbf{X}'\mathbf{F}
\mathbf{y}.
\]
Indeed, setting $\mathbf{c}=\mathbf{X}_{b}'\mathbf{F}_{b}\mathbf{y}$,
the first coordinate of the solution vector is
\[
\hat{\beta}_{0}=\frac{\det(\tilde{\mathbf{H}}_{4}^{0}[1, \mathbf
{c}])}{\det(\mathbf{H}_{s})}=\frac{\det(\mathbf{H}_{4}^{0}[1, \mathbf
{c}])}{\det(\mathbf{H}_{s})}.
\]
Given that $\mathbf{c}=\sum_{j=-m}^{m}(\mathbf
{j/b})(1/b)f_{0B}(j/b)y_{t+j}$, it follows that
\[
\det\bigl(\mathbf{H}_{4}^{0}[1, \mathbf{b}]\bigr)=\sum
_{j=-m}^{m}\det\biggl(\mathbf
{H}_{4}^{0}\biggl[1,\frac{\mathbf{j}}{\mathbf{b}}\biggr]\biggr)
\frac{1}{b}f_{0B} \biggl(\frac
{j}{b} \biggr)y_{t+j}
\]
and, therefore,
\[
\hat{g}_{t}=\sum_{j=-m}^{m}
\frac{\det(\mathbf{H}_{4}^{0}[1, \mathbf
{j}/\mathbf{b}])({1}/{b})f_{0B} ({j}/{b} )}{\det(\mathbf{H}_{s})}y_{t+j}.
\]
Hence,
\[
\hat{\beta}_{0}=\mathbf{e}_{1}'{
\mathbf{H}_{s}}^{-1}\mathbf {X}_{b}'
\mathbf{F}_{b}\mathbf{y},
\]
and it follows that
\[
\mathbf{w}'=\mathbf{e}_{1}'{
\mathbf{H}_{s}}^{-1}\mathbf{X}_{b}'
\mathbf{F}_{b}.
\]
\end{appendix}

\section*{Acknowledgments}
We are indebted to the Editor Professor Brendan Murphy for his valuable
comments that helped to enrich a previous version of this manuscript.
We also want to thank the Associate Editor and two referees for their
insightful suggestions and comments.

%\vfill
%\vfill
%\begin{appendix}
%\section{}
%\end{appendix}

% zodis "Acknowledgments" paliekamas pagal autoriu
%\section*{Acknowledgments}

\begin{supplement}%[id=suppA]
%\sname{Supplement A}
\stitle{Weight systems}
\slink[doi]{10.1214/15-AOAS856SUPP} %[doi,text={...}] - jei reikia
%suskaldyti doi
\sdatatype{.pdf}
\sfilename{aoas856\_supp.pdf}
\sdescription{The supplementary material contains the weight systems of
our filters for 9- and 13-term symmetric filters.}
\end{supplement}

%\newpage
%
% imsref loaded by akundreckaite, 2015-08-21 08:46:24
% imsref loaded by akundreckaite, 2015-08-21 08:55:09
% imsref loaded by akundreckaite, 2015-08-21 12:58:20
% imsref loaded by akundreckaite, 2015-08-24 10:19:33

%\begin{thebibliography}{99}
%\bibitem{r1}
%\bibitem{r1}
%\end{thebibliography}

\printaddresses
\end{document}